\documentclass[apj]{emulateapj}
\usepackage{natbib}

\usepackage{amsmath}
\usepackage{apjfonts}
\usepackage{graphicx}
\usepackage{amssymb}
\shorttitle{Probing Asymmetric Structures in the outskirts of Galaxies}
\shortauthors{Wen, Zheng \& An}

\begin{document}

\title{Probing Asymmetric Structures in the outskirts of Galaxies}
\author{Zhang~Zheng~Wen\altaffilmark{1,2}, Xian~Zhong~Zheng\altaffilmark{1} and Fang~Xia~An\altaffilmark{1}}
\altaffiltext{1}{Purple Mountain Observatory, Chinese Academy of Sciences, 2 West Beijing Road, Nanjing, 210008, China}
\altaffiltext{2}{University of Chinese Academy of Sciences, 19A Yuquan Road Beijing, China}

\begin{abstract}
Upcoming large imaging surveys will allow detailed studies of the structure and morphology of galaxies aimed at addressing how galaxies form and evolve. Computational approaches are needed to characterize their morphologies over large samples. We introduce an automatic method to quantify the outer structure of galaxies. The key to our approach is the division of a galaxy image into two sections delineated by the isophote which encloses half the total brightness of the galaxy. We call the central section the inner half-flux region (IHR) and the outer section the outer half-flux region (OHR). From this division, we derive two parameters: $A_{\rm o}$, which measures the asymmetry of the OHR, and $D_{\rm o}$, which measures the deviation of the intensity weighted centroid of the OHR from that of the IHR relative to the effective radius. We derive the two parameters from $HST$/ACS $z_{850}$-band images for a sample of 764 galaxies with $z_{850}<22$\,mag and $0.35<z<0.9$ selected from GEMS and GOODS-South surveys. We show that the sample galaxies having strong asymmetric structures, in particular tidal tails, are well-separated from those with regular morphologies in the $A_{\rm o}$-$D_{\rm o}$ space. Meanwhile, the widely used {\it CAS} and Gini-$M_{20}$ methods turn out to be insensitive to such morphological features. We stress that the $A_{\rm o}$-$D_{\rm o}$ method is an efficient way to select galaxies with significant asymmetric features like tidal tails and study galaxy mergers in the dynamical phase traced by these delicate features.
 \end{abstract}

\section{INTRODUCTION}

       Major mergers between galaxies of comparable mass are expected to occur frequently in hierarchical models of galaxy formation and evolution \citep[e.g.,][]{1978MNRAS.183..341W,1993MNRAS.262..627L}. Galaxy merging may be a crucial process that regulates galaxy mass assembly, galaxy morphology reshaping, growth of supermassive black holes, and enhancement of star formation \citep[e.g.,][]{1992ARA&A..30..705B,1996ARA&A..34..749S,2005Natur.435..629S,2008ApJS..175..356H,2010ApJ...724..915H,2009ApJ...697.1369B,2014arXiv1403.2783C}. Measuring the galaxy merger rate over cosmic time is thus a central task in determining the importance of the merging process relative to that of other physical processes (e.g., feedback and gas accretion) to driving galaxy evolution. 

Much effort has been made to measure galaxy merger rate. Interacting or merging galaxies are rare ($\sim$2\%) in the local universe \citep{1985ARA&A..23..147A,1997ApJ...475...29P,2004ApJ...603L..73X,2008ApJ...685..235P,2009ApJ...695.1559D}, but they become more numerous at higher redshifts. However, the measurements of the merger rate are inconsistent with each other and are still under debate. While strong evolution characterized by $(1 + z)^{3-6}$ was often reported \citep{2000MNRAS.311..565L,2002ApJ...565..208P,2003AJ....126.1183C,2005MNRAS.357..903C,2007ApJS..172..329K,2007ApJS..172..320K,2009MNRAS.394.1956C,2009ApJ...697.1971J,2010ApJ...713..330X,2011ApJ...742..103L}, mild evolution following $\sim (1+z)^{0.5}$ or even no evolution was obtained by other studies \citep{2000ApJ...532L...1C,2004ApJ...601L.123B,2008ApJ...681..232L,2010ApJ...719..844R,2012ApJ...744...85M}. The controversy is likely caused by large uncertainties in current observational techniques adopted for merger identification (see \citealt{2008MNRAS.391.1137L} for more details).

Disturbed morphology is mostly used as the probe of a merger. 
The violent tidal forces between merging galaxies can destroy galaxy structures and produce tidal features. For instance, extended tidal tails can be created when the mergers involve a disk galaxy \citep{1972ApJ...178..623T,1972MNRAS.157..309W,1992Natur.360..715B,2004IAUS..217..390M}. The characteristics of the tidal tails in turn disclose key properties of the parent galaxies such as kinematics, mass ratio, and orbital parameters \citep[][]{2013ASPC..477...47D,2013LNP...861..327D}. A long tidal tail is usually seen as evidence for a merger between disk galaxies \citep[e.g.,][]{2007ApJ...663..734E,2010ApJ...709.1067B}. Resolving specific features from mergers helps to provide information on the frequency of mergers, a better understanding of the merger time scale, and a complete census of various types of mergers over cosmic time. It further delivers insights on evolutionary pathways for different galaxy populations. This will be possible with upcoming deep imaging surveys over large areas \citep[e.g., Euclid,][]{2013LRR....16....6A}. Computational approaches are keenly needed to detect the specific features tracing given types of mergers.

Non-parametric methods $CAS$ \citep{2000ApJ...529..886C, 2003ApJS..147....1C} and Gini-$M_{20}$ \citep{2004AJ....128..163L} are widely used for merger selection.  $CAS$ selects mergers mainly according to their morphological asymmetry, but fails to pick up those with weaker asymmetry in morphology (e.g., double-nucleus). The Gini coefficient and $M_{20}$ parameters measure the relative distribution of pixel fluxes and spatial light concentration of a galaxy, respectively. Mergers and regular galaxies are globally separated in the Gini-$M_{20}$ space. This method favors to select mergers that are in their first pass or the final stage \citep{2008MNRAS.391.1137L,2010MNRAS.404..575L,2010MNRAS.404..590L}. Neither $CAS$ nor Gini-$M_{20}$ provides a complete selection for major mergers \citep{2007ApJS..172..329K,2007ApJS..172..406S,2009ApJ...697.1971J,2010ApJ...721...98K}. This is in part because the parameters adopted in the two methods are flux weighted. These parameters are largely determined by the light distribution of the bright section of a galaxy, and are insensitive to the tidal features with lower surface brightness in the outskirts of the galaxy.

In this paper, we present a new approach to quantifying the structure of the outskirts of a galaxy, aimed at searching for tidal tails. Two parameters are introduced to accomplish this. In Section~\ref{sec:sec2} we describe how to measure the two parameters. In Section~\ref{sec:sec3}, we verify our method  using a visually classified sample of galaxies. Finally, we compare our method with $CAS$ and Gini-$M_{20}$ in Section~\ref{sec:sec4}. Throughout this paper, we assume $H_0=70$\,km\,s$^{-1}$\,Mpc$^{-1}$, $\Omega_{\rm \Lambda}=0.7$ and $\Omega_{\rm m}=0.3$.

\section{METHODOLOGY} \label{sec:sec2}

\subsection{Galaxy Division}\label{sec:sec2.1}

    We divide the image of a galaxy into two sections split by the isophote which encloses half the total light of the galaxy. The inner section, named as the inner half-flux region (IHR), usually encloses a smaller area with a higher surface brightness than the outer section, named as the outer half-flux region (OHR). In practice, the IHR is identified first, and the rest of the galaxy is presumed to occupy the OHR. If the galaxy is regular in morphology, the two sections can be simply separated by the isophote which contains half the total light. If a galaxy does not have a single, clear core, but has double or even multiple comparable components, then we fix the center at the brightest component, and find the half-light isophote around this point.

   We develop a technique to deal with image pixels and identify the brightest group of contiguous pixels as the brightest component in a galaxy image. 
The software tool SExtractor \citep{1996A&AS..117..393B} is used to detect pixels that a galaxy covers in the background-subtracted image (see Section~\ref{sec:sec3.3} for details). These pixels are sorted in descending order of flux. The pixels with highest fluxes are selected, yielding $f$, a ratio of the integrated flux to the total. More pixels within the galaxy will be included if a higher $f$ is set.  Starting from $f$=0.5, the selected pixels may spread into two or more discrete groups of contiguous pixels for galaxies with complex light distribution (e.g., late-type spirals with giant clumps). 
These discrete groups of pixels are then individually examined to obtain their integrated fluxes. By increasing $f$, each group is expected to contain more contiguous pixels and thus a higher fraction of the total flux. The brightest group with a sum of fluxes higher than 25\% of the total is chosen to calculate a flux-weighted centroid. This position is taken as the center for isophotal fitting to the galaxy image. 
For a galaxy with smooth light distribution usually satisfying a S\'{e}rsic function, the selected pixels make up one group, which is indeed the IHR when $f=0.5$. For a merging system with two nuclei, the brighter one will be selected. The factor of 25\% is decided by this selection.

\begin{figure}[]
  \centering
    \includegraphics[width=0.47\textwidth]{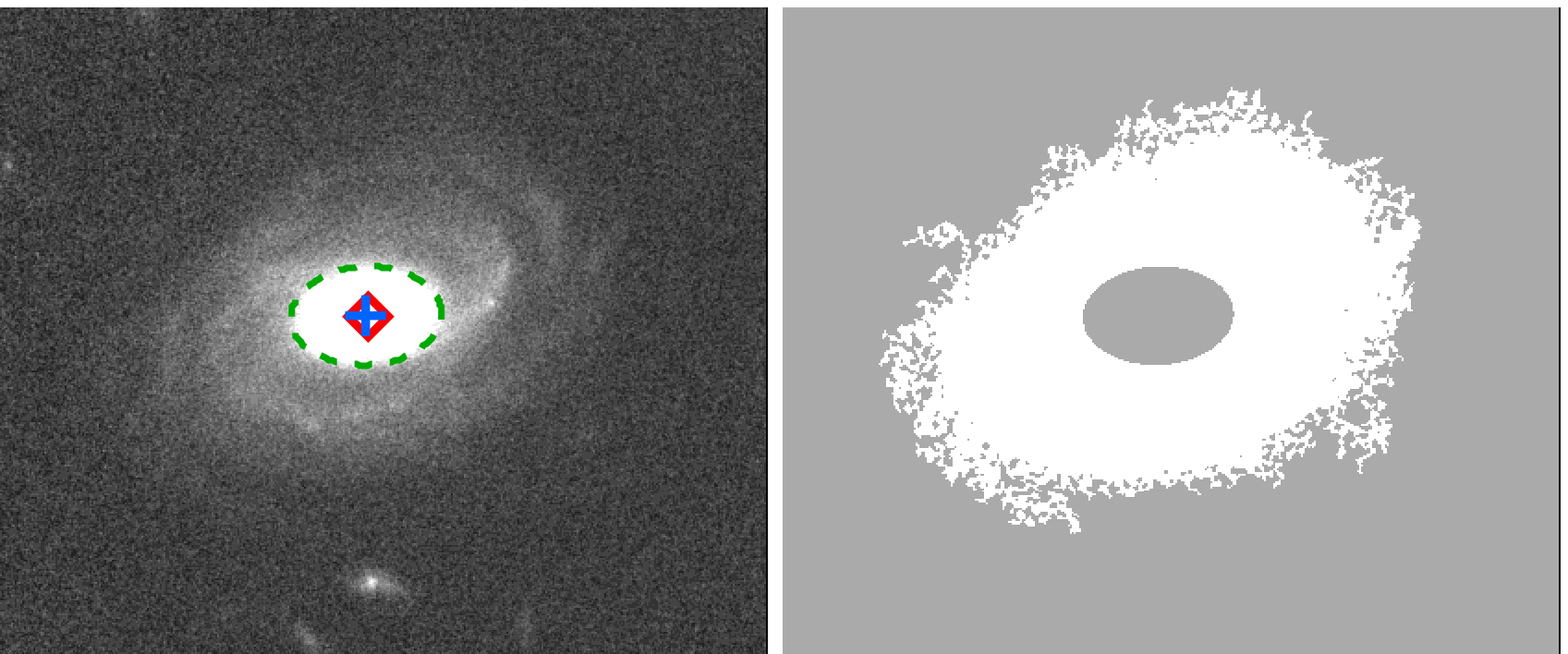}
    \includegraphics[width=0.47\textwidth]{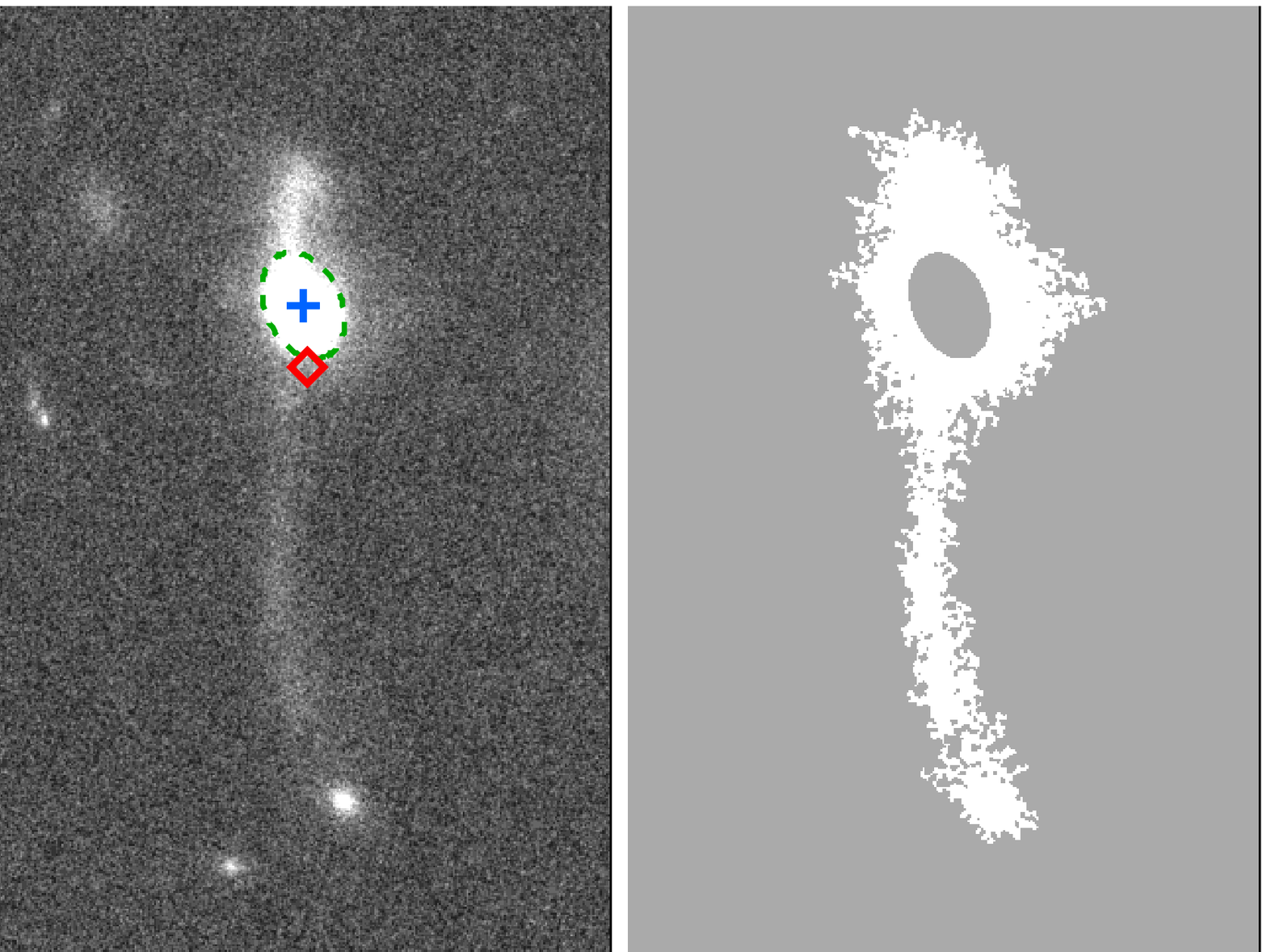}
  \caption{$HST$/ACS $z_{850}$-band images (left) and segmentation maps (right) for a typical spiral galaxy at $z=0.47$ (top) and a merger with tidal tail at $z=0.58$ (bottom). The ellipses (dashed lines) mark the IHR. The blue crosses and red diamonds give the centroids of the IHR and OHR of each galaxy, respectively. }
\label{fig:fig1}
\end{figure}

  With the center fixed, elliptical isophotes are fitted to the galaxy image. Following the algorithms given in SExtractor, the ellipticity and position angle of an isophote are determined by the maximum and minimum spatial dispersion of the enclosed pixels by the real isophote. The ellipse containing 50\% of the total flux is taken as the boundary for the IHR. And the pixels out of the isophote make up the OHR. Figure~\ref{fig:fig1} demonstrates the separation between the IHR and OHR for two representative galaxies.

\subsection{Outer asymmetry} \label{sec:sec2.2}
     
The asymmetry parameter ($A$) is often used to quantify the morphological disturbance of a galaxy \citep{1996MNRAS.279L..47A,2000ApJ...529..886C,2003ApJS..147....1C}. It is calculated as the sum of the absolute residuals of the galaxy image subtracting itself rotated by 180$\arcdeg$ around its center. The rotation center is determined iteratively in order to minimize the asymmetry. We define outer asymmetry, $A_{\rm o}$, as the asymmetry of the OHR of a galaxy following 

\begin{equation} \label{equ:eq1}
A_{\rm o} =\frac{\sum{|I_{\rm o}-I_{\rm o}^{180^\circ}|}}{\sum{I_{\rm o}}}-\frac{\sum{|B_{\rm o}-B_{\rm o}^{180^\circ}|}}{\sum{I_{\rm o}}}, 
\end{equation}

where $I_{\rm o}$ is flux distribution of the OHR and $I_{\rm o}^{180^\circ}$ represents the $180^\circ $-rotated $I_{\rm o}$. In the same way, $B_{\rm o}$ is a patch in the background with the same shape as $I_{\rm o}$. Again, $B_{\rm o}^{180^\circ}$ is the $180^\circ $-rotated $B_{\rm o}$.

The rotation center is critical to the estimation of the asymmetry parameter. For very nearby galaxies, a small shift for the center would cause a significant change in the estimated asymmetry. For distant galaxies, determining a rotation center becomes less complicated because of decreasing resolution \citep{2000ApJ...529..886C}. We thus adopt the centroid of the OHR as the rotation center when estimating $A_{\rm o}$.

\subsection{centroid deviation}\label{sec:sec2.3}

The centroid deviation parameter, $D_{\rm o}$, measures the deviation (or offset) between centroids of the IHR and OHR of a galaxy.  The centroid is determined by
\begin{equation} \label{equ:eq2}
x_{\rm cen} = \frac{\sum\limits_{i\in S} f_i~ x_i }{ \sum\limits_{i\in S} f_i } 
~~ {\rm and}  ~~
y_{\rm cen} = \frac{\sum\limits_{i\in S} f_i~ y_i }{ \sum\limits_{i\in S} f_i },
\end{equation}
where $x_i$ and $y_i$ define the position of pixel $i$ in the galaxy image and $f_{\rm i}$ is flux of the pixel.  
The centroid deviation is calculated using the formula 
\begin{equation} \label{equ:eq3}
D_{\rm o}=\frac{\sqrt{(x_{\rm o}-x_{\rm c})^2+(y_{\rm o}-y_{\rm c})^2}}{R_{\rm e}}, 
\end{equation}
where ($x_{\rm c}$,$y_{\rm c}$) and ($x_{\rm o}$,$y_{\rm o}$) refer to the centroids of the IHR and OHR, respectively. The effective radius of the galaxy, $R_{\rm e}$, is used for normalization. Here $R_{\rm e}$ is estimated using $R_{\rm e}=\sqrt{n_{\rm c}/\pi}$, where $n_{\rm c}$ is pixel count of the IHR.

For regular galaxies with symmetric morphologies, the IHR and OHR share nearly the same centroid, yielding a centroid deviation near zero. For merging galaxies with tidal tails, however, the centroid deviation becomes rather large, as shown in Figure~\ref{fig:fig1}. Similarly, the merging galaxies tend to have higher $A_{\rm o}$ than the regular galaxies. Therefore, galaxies with asymmetric structures are expected to be away from the regular galaxies in the diagram of $A_{\rm o}$ versus $D_{\rm o}$. We test this $A_{\rm o}$-$D_{\rm o}$ method using a sample of galaxies with high-resolution images from the Hubble Space Telescope ($HST$).

\section{VERIFYING THE $A_{\rm o}$-$D_{\rm o}$ METHOD} \label{sec:sec3}
  
\subsection{Galaxy sample} \label{sec:sec3.1}

The extended Chandra Deep Field South (ECDFS) is one of the cosmic fields with the deepest multi-wavelength data over an area of $30\arcmin \times 30\arcmin$, including $HST$/ACS imaging data from the GEMS \citep{2004ApJS..152..163R} and the GOODS-S surveys \citep{2004ApJ...600L..93G}, the survey catalog \citep{2008ApJS..174..136C},  photometry and photometric redshifts from the Multiwavelength Survey by Yale-Chile \citep[MUSYC;][]{2010ApJS..189..270C}, and the stellar mass catalog from the COMBO-17 survey \citep{2006A&A...453..869B}. The best-fit models of two-dimensional S\'{e}rsic profile derived from $HST$ images are available for GEMS galaxies \citep{2007ApJS..172..615H}. We use these data to select a sample of galaxies and test our method for quantifying morphologies of the OHRs.

We focus on massive galaxies  that enable strong tidal features to be produced in the merging process and detectable at high-$z$. We derive morphological parameters from $HST$/ACS images in the $z_{850}$ band, which corresponds to the rest-frame $B$ to $V$ in the redshift range $0.35<z<0.9$. In total, 825 galaxies are selected with stellar mass $\log (M/{\rm M}_\odot) \geq 10.5$ and $0.35<z<0.9$ over the 800 square arcminutes area covered by $HST$ observations. Of them, 32 objects are too compact to be resolved in morphology; and 29 objects are located at image edges. Finally, 764 galaxies are selected to compose a mass-complete sample for our morphological investigation. The sample galaxies are mostly with $z_{850}$-band magnitude $\sim 20 - 22$, compared to the 5\,$\sigma$ depth of $z_{850}=27.1$\,mag.

The S\'{e}rsic index ($n$) is often used as a morphology indicator (e.g., $n<$2.5 for disk galaxies). However,  the two-dimensional structural fitting ignores irregular structures that are crucial to identifying galaxy mergers, although the S\'{e}rsic models are good proxies for regular galaxies. Visual examination is still an efficient way to ascertain the tidal features. Our $A_{\rm o}$-$D_{\rm o}$ method lacks sensitivity to distinguish galaxies with regular morphologies. We therefore adopt seven morphological types optimized for verifying the $A_{\rm o}$-$D_{\rm o}$ method, including spheroids (E, S0), early disks (Sa-b), late disks (Sc-d), edge-on disks, irregulars/minor mergers, major mergers without tidal tail, and major mergers with tidal tails. The morphological classification was independently carried out by all three co-authors. As a result, the 764 galaxies in our sample are classified into  440, 57, 57, 27, 108, 60 and 15 from the first to the last class, respectively. 
We are interested in the latter three morphological classes, in particular major mergers with tidal tails.  We note that tidal tails with a size of tens of kpcs at intermediate redshifts can be resolved with $HST$/ACS imaging. 
We explain in detail how to recognize a tidal feature as a tidal tail.

\begin{figure*}[] 
\centering
\includegraphics[width=\textwidth]{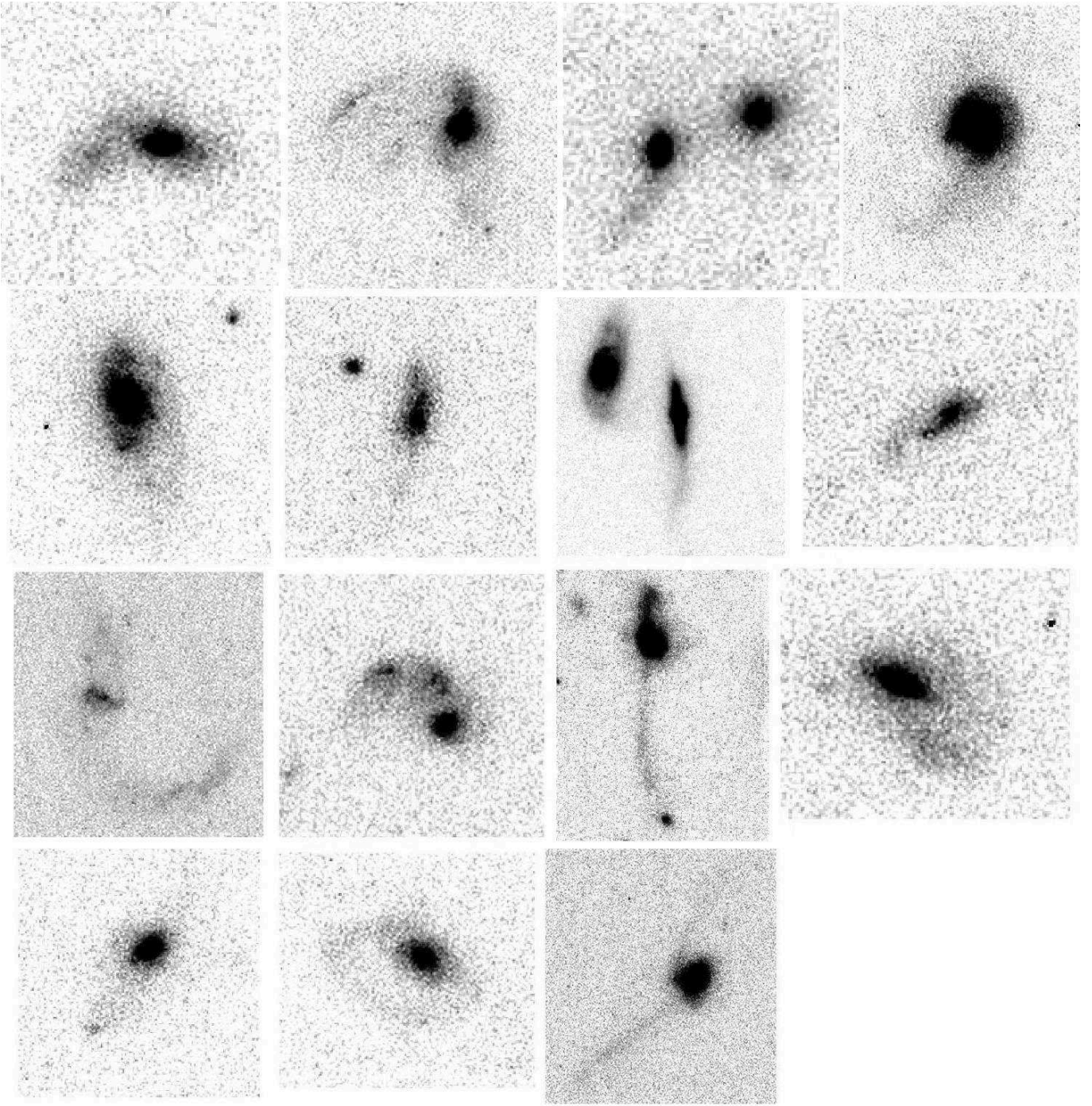}
\caption{A sample of mergers with tidal tails in the ECDFS field. 
From left to right, galaxy IDs are 6494, 7357, 11072, 17207 (top), 20158, 24090, 29976, 30004 (the 2nd row), 30076, 44488, 45115, 57822 (the 3rd row), 57896, 60651, and 61546 (bottom).}
\label{fig:fig2}
\end{figure*}

\begin{figure*}[] 
\centering
\includegraphics[width=0.9\textwidth]{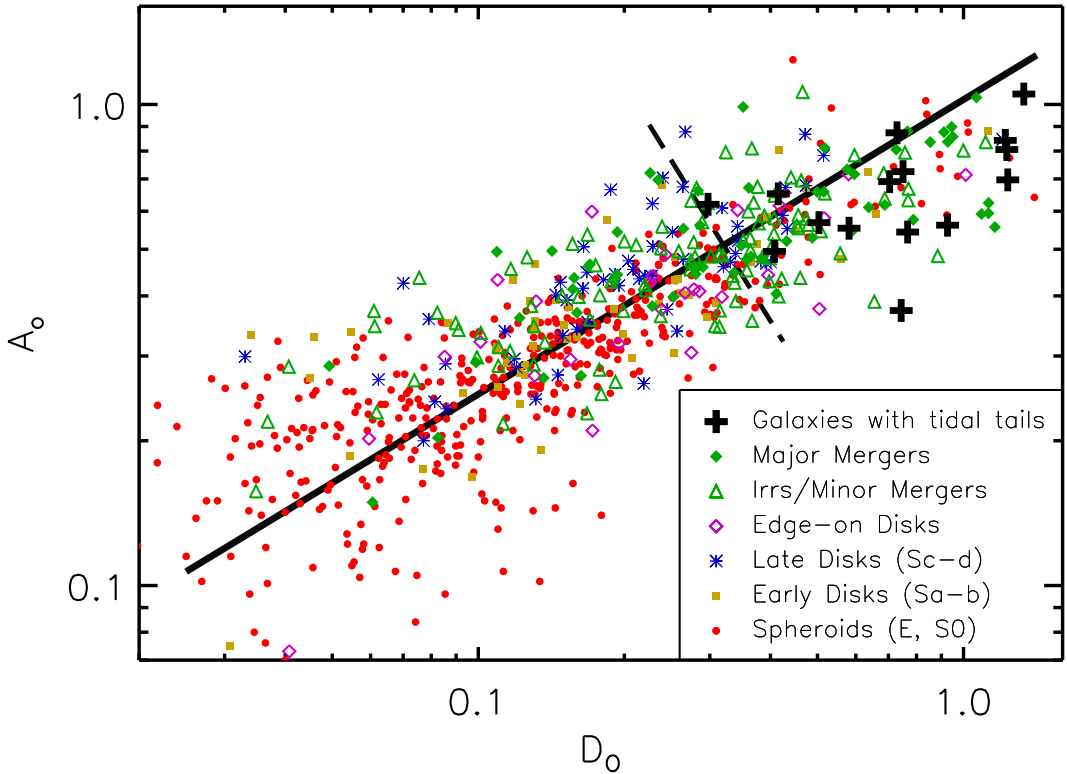}
\caption{The $A_{\rm o}$-$D_{\rm o}$ diagram of our sample of 764 galaxies with $\log (M/\rm M_\odot) \geq 10.5$ and $0.35<z<0.9$ in the ECDFS. The two parameters are derived from {\it HST} $z_{850}$-band images, corresponding to the rest-frame optical for the redshift range examined here.  The dashed line is the criterion for selecting tidal tails described as ${\rm log}\,A_{\rm o} > -1.6\,{\rm log}\,D_{\rm o}-1.1$. The solid line represents the sequence of regular galaxies best-fitted by ${\rm log}\,A_{\rm o}=0.6\,{\rm log}\,D_{\rm o}$.}
\label{fig:fig3}
\end{figure*}

\begin{figure*}[] 
\centering
\includegraphics[width=\textwidth]{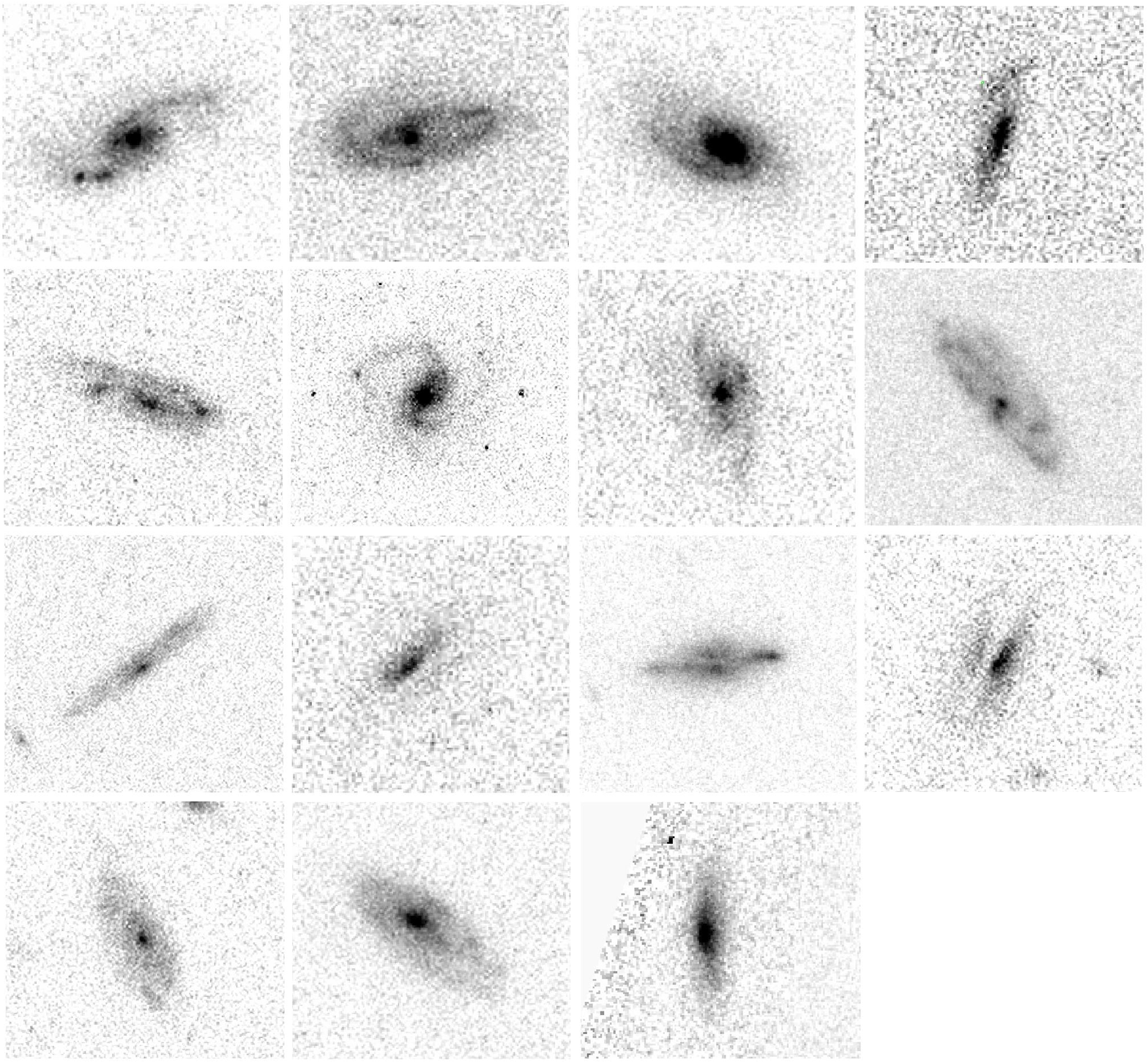}
\caption{{\it HST} $z_{850}$-band images of sample galaxies satisfying the criterion ${\rm log}\,A_{\rm o} > -1.6\,{\rm log}\,D_{\rm o}-1.1$ and $D_{\rm o}<0.5$. These galaxies are visually classified as spheroids or disks. From left to right, the galaxy IDs are: 49368,  1032,  29488, 35796 (top); 61789,  2848,  42774,  42874 (the 2nd row); 20664,  30167,  30189,  56918 (the 3rd row); 37335,  59517, and 51329 (bottom).}
\label{fig:fig4}
\end{figure*}

\subsection{The properties of tidal tails} \label{sec:sec3.2}
    
    The visual investigation by \citet{2007ApJ...663..734E} revealed that the host merging galaxies of tidal tails in ECDFS can be divided into four morphological types: diffuse, antennae, M51 and shrimp. Figure~\ref{fig:fig2} shows $HST$ images of merging galaxies with tidal tails selected from the ECDFS field. 

The diffuse-type galaxies refer to those with diffuse arc-like or tail-like intergalactic structures \citep[e.g.,][]{2006ApJ...648..969K}. Such tidal structures tend to have red color and smooth light profile, suggesting that they are tidal debris of early-type galaxies with little gas through dry mergers (e.g., the top-right panel in Figure~\ref{fig:fig2}). 

   The antennae-type galaxies have disturbed merger cores and extended tidal tails. They are most likely formed by major mergers between disk galaxies \citep[e.g.,][]{1972ApJ...178..623T}. 
The visible timescale for tidal tails can be a few hundred Myrs \citep{2009MNRAS.399L..16C} up to a few Gyrs \citep{1995AJ....110..140H}. About half the merging galaxies with tidal tails shown in Figure~\ref{fig:fig2} are antennae-type (e.g. ID 7357, 44488, 45115, 57896).
Tidal dwarf galaxies can be formed in these tails, being a result of local gravitational instabilities of the gas component \citep{1993ApJ...412...90E,2006A&A...456..481B,2007MNRAS.375..805W} or at the tip of a tidal tail regulated by the accumulation and collapse of massive gaseous condensations \citep{2004A&A...427..803D}. The presence of long tidal tails or tidal dwarfs dramatically increases $A_{\rm o}$ and $D_{\rm o}$ (see Section~\ref{sec:sec3.3}).

     The M51-type galaxies are mergers at their early stages and the stellar bridge connecting one galaxy to the another is still seen. The galaxies of this type are rare probably because of the short timescale to be detected \citep[several Myrs; ][]{2006A&A...456..481B}.
Shown in Figure~\ref{fig:fig2}, 11072 is a pair of galaxies of comparable $z_{850}$-band luminosity. The system has just undergone its first pericenter passage with a weak tidal bridge. A tail-like structure is seen with color $V_{606}-z_{850}=1.28$, similar to the color of the central components ($V_{606}-z_{850}=1.33$). No stellar clump is found on the tidal tail. 

 Finally, shrimp-type galaxies are characterized by a highly warped, dominant arm or tail and have no well-defined central nucleus. This type of merger is not included in Figure~\ref{fig:fig2}.

\subsection{The $A_{\rm o}$-$D_{\rm o}$ classification} \label{sec:sec3.3}

Using $HST$ images, we calculate outer asymmetry $A_{\rm o}$ and centroid deviation $D_{\rm o}$ for our sample of 764 galaxies. We run SExtractor to create a segmentation map for every galaxy. The segmentation map is an image with the same size as the scientific image where objects and their boundaries are determined. The total flux of a galaxy is then defined as the light contained in the pixels within its boundary. A detection threshold of 0.8\,$\sigma_{\rm bkg}$ above background is chosen. Here $\sigma_{\rm bkg}$ is the $RMS$ of background noise. Other configuration parameters are set following \citet{2008ApJS..174..136C} for the GEMS catalog. This threshold is lower than usual settings \cite[e.g., 1.65\,$\sigma_{\rm bkg}$,][]{2008ApJS..174..136C} for source detection, aimed at finding extended tidal features that tend to have lower surface brightness at larger radii of the OHR of a galaxy. We point out that the total flux adopted here differs from that given, for example, in the Third Reference Catalogue of Bright Galaxies \citep[RC3;][]{1991rc3..book.....D}, where missing flux is added in by extrapolation of integrated luminosity profiles to infinite radius. This procedure is impractical for our analysis since we are not fitting light profiles. The measurements of $A_{\rm o}$ and $D_{\rm o}$ are nevertheless likely to be only marginally affected by this difference.

We test the $A_{\rm o}$-$D_{\rm o}$ method for quantifying galaxy morphologies using our sample of 764 galaxies in comparison with the morphologies visually classified. Figure~\ref{fig:fig3} illustrates the distribution of the sample in $A_{\rm o}$-$D_{\rm o}$ diagram. Different morphological types are marked with different symbols. We can see that all sample galaxies lie on a tight sequence, where the location is in general correlated with the degree of morphological disturbance. Galaxies with stronger disturbed morphologies tend to have higher $A_{\rm o}$ and $D_{\rm o}$. The vast majority of galaxies with regular morphologies (E, S0, Sa-b, Sc-d, edge-on disks) have $D_{\rm o}<0.5$ and $A_{\rm o}<0.6$, following a tight relation ${\rm log}\,A_{\rm o}=0.6\,{\rm log}\,D_{\rm o}$. 
This relation indicates that the two parameters $A_{\rm o}$ and $D_{\rm o}$ are correlated with each other.  This is also confirmed by the fact that no data points are found in the bottom-right and top-left regions of the diagram. 

However, different morphological types of the sample galaxies are not clearly separated along the sequence. 
Spheroids and early disks having symmetric morphologies reside mostly in the region $D_{\rm o}<0.2$. Edge-on disks and late disks with non-symmetric substructures like arms, H\,{\footnotesize II} regions and stellar clusters are distributed more widely over the sequence. Such substructures are unlikely to be responsible for a high $D_{\rm o}$. Irregular galaxies and major/minor mergers are systematically higher in $A_{\rm o}$ but spread over the full range of $D_{\rm o}$. 

The striking result from Figure~\ref{fig:fig3} is that all galaxies with tidal tails (black crosses) have apparently higher $D_{\rm o}$ and higher $A_{\rm o}$, compared with the galaxies of other morphological types, in particular regular ones (spheroids and disks). The large spread in $D_{\rm o}$ is mostly caused by the variety of the tidal tails in size and light weight relative to their host galaxies. The majority (12/15) of these with tidal tails are at $D_{\rm o}>0.5$.
This indicates that $D_{\rm o}$ is sensitive to the tidal tails, although three objects with weak tidal tails (17207, 20158, 24090) have $D_{\rm o}<0.5$.

As shown in Figure~\ref{fig:fig3}, the criterion ${\rm log}\,A_{\rm o} > -1.6\,{\rm log}\,D_{\rm o}-1.1$ is sufficient to select all galaxies with tidal tails and most major mergers (32/60) in our sample. This cut, however, also picks up a large fraction (43\%) of objects with irregular morphologies (irregulars/minor mergers).
We note that some regular galaxies (spheroids and disks) also fall into the $D_{\rm o}>0.5$ region. We examined their images, and found that their $A_{\rm o}$ and $D_{\rm o}$ values are significantly affected by the unmasked residual light from neighboring sources. There are also 45 regular galaxies satisfying ${\rm log}\,A_{\rm o} > -1.6\,{\rm log}\,D_{\rm o}-1.1$ and $D_{\rm o}<0.5$. Again, we visually checked their $HST$ images and segmentation maps. Of the 45 targets,  23 are contaminated by neighboring sources, and another 22 are isolated with asymmetric outskirts. Again the two morphological parameters $A_{\rm o}$ and $D_{\rm o}$ are largely overestimated for the visually-classified regular galaxies (spheroids and disks) due to the contamination. 
Figure~\ref{fig:fig4} shows the 15 regular galaxies meeting the selection.
We further investigate their morphological properties in detail for a better understanding of the selection.

\begin{itemize}
\item[] {\it Galaxy 49368}: it has bright stellar clumps of 1\,kpc size on one spiral arm, which seriously bias the estimation of $A_{\rm o}$ and $D_{\rm o}$. Such features can be frequently seen in $z>2$ disk galaxies. Minor mergers and instabilities in gas rich, turbulent disks are likely responsible for the formation of such clumps \citep{2008ApJ...687...59G,2009ApJ...694L.158B,2009ApJ...703..785D}.

\item[] {\it Galaxies 1032, 29488, 35796 and 61789}: they are characterized by one prominent spiral arm.  Their morphologies are similar to the so-called shrimp-type galaxies defined by \citet{2007ApJ...663..734E}. The highly curved spiral arm together with star-forming clumps on it usually imply that they are gas-rich systems disturbed by minor mergers. However, no apparent signature is found for tidal disturbance in the four galaxies. 

\item[] {\it Galaxies 2848, 42774 and 42874}: they have multiple spiral arms and show strong asymmetry in morphology.

\item[] {\it Galaxies 20664, 30167, 30189 and 56918}: they are edge-on disk galaxies with a clear dust lane feature. The dust lane divides the light profile of the parent galaxy into two parts. In such cases, the system is treated as a double core system in our analysis. The centroid of the IHR would significantly deviate from the geometric center if the dust lane is thick and splits the galaxy into two non-equal parts.

\item[] {\it Galaxies 37335 and 59517}: they are classified as Sd galaxies. Their surface brightness is relatively low. And their $A_{\rm o}$ and $D_{\rm o}$ are largely biased by clumpy light distribution and background noise.

\item[] {\it Galaxy 51329}: it is an edge-on disk galaxy. The disk is apparently warped.  Minor mergers/interactions are often proposed as the cause of warped disks \citep{1990MNRAS.246..458S,1991A&A...242..301F,1998A&A...337....9R,2006NewA...11..293A}.
\end{itemize}

It is worth noting that the segmentation strategy used in our analysis relies on a low detection threshold. This may potentially mistake some noise pixels for part of the target, resulting in a fuzzy appearance at edges. Since outer asymmetry and centroid deviation measure the morphological distribution of the OHR, the background noise may have noticeable effect on the estimate of the two parameters, in particular for low surface brightness galaxies.

  Source deblending is also one issue that may affect the measurements of $A_{\rm o}$ and $D_{\rm o}$. This is also key to extracting source catalogs from deep images.  We adopted the optimal configuration for Sextractor from the GEMS catalog to produce segmentation maps for the selected sample galaxies. While the vast majority of sources can be properly resolved in the $HST$ images with the configuration (see \citealt{2008ApJS..174..136C} for more details), a small fraction ($<\sim$5\%) of sources may still be contaminated by foreground or background sources due to projection effects, or over deblended into multiple sources. Merging galaxies with disturbed morphologies tend to suffer more from this problem. In this work, we visually examined the deblending for every sample galaxy and found a few cases with contamination by the neighboring sources which mislead the targets towards a high outer asymmetry.

\begin{figure}[] 
\centering
\includegraphics[width=0.455\textwidth]{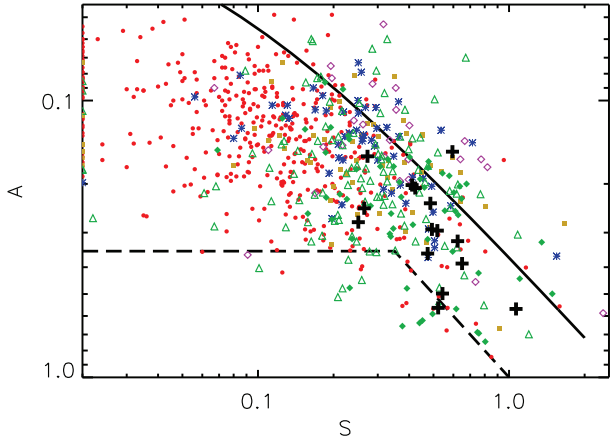}
\caption{The relation between asymmetry ($A$) and clumpiness ($S$) for our sample. The solid line $A=0.35\,S+0.02$ gives the relationship of nearby galaxies that are not involved in mergers while the dashed lines are the criteria ($A>0.35, A>S$) for selecting merging galaxies \citep{2003ApJS..147....1C}. The data points are given in the same way as in Figure~\ref{fig:fig3}.}
\label{fig:fig5}
\end{figure}

 \section{COMPARISON WITH  $CAS$ AND Gini-$M_{20}$} \label{sec:sec4}

The non-parametric methods $CAS$ and Gini-$M_{20}$ are widely used automatic approaches in the literature to identify merging galaxies. In this section, we compare our $A_{\rm o}$-$D_{\rm o}$ method with the $CAS$ and Gini-$M_{20}$ methods in finding galaxies with extended tidal features. 

\subsection{$CAS$}   \label{sec:sec4.1}

The $CAS$ method \citep{2000ApJ...529..886C,2003ApJS..147....1C} involves three morphological parameters: concentration ($C$), asymmetry ($A$) and clumpiness (or smoothness; $S$). The $C$ parameter describes the intensity of stellar light contained within the central region in comparison to a larger outer region of a galaxy. The $A$ quantifies the degree that the surface brightness profile of the galaxy deviates from a perfectly symmetric distribution. The $S$ parameter focuses on the flux that is contained in the clumpy part of the light distribution. We calculate $A$ and $S$ parameters for our sample galaxies using {\it HST} $z_{850}$-band images. The results are shown in Figure~\ref{fig:fig5}. The solid line represents the $S-A$ relationship for regular galaxies in the local universe \citep{2003ApJS..147....1C}. It has been shown that the star formation activities lead to an increase in both $S$ and $A$ for nearby galaxies. We find, however, that our sample galaxies at intermediate redshifts deviate from the relationship of nearby galaxies. This has also been reported and explained in \citet{2008MNRAS.386..909C}. The reason for such a deviation is that both $S$ and $A$ decrease when the resolution and signal-to-noise ratio (S/N) become lower. And $S$ is affected by a larger degree than $A$. 

   In $CAS$, a galaxy is identified as a major merger when $A>0.35$ and $A>S$. This selection works well for those where morphological distortion affects more than 35\% of the total light of the galaxy. The condition $A>S$ ensures that the asymmetric light is not dominated by clumpy star-forming regions. Figure~\ref{fig:fig5} shows that only about one quarter (16/60) of the major mergers (green diamonds) and  one third (5/15) of galaxies with tidal tails (black crosses) in our sample satisfy $A>0.35$.
This is not surprising, as $A>0.35$ statistically accounts for one third of the timescale for a major merger determined by N-body simulations, while minor mergers with mass ratio below $1:5$ hardly reach $A>0.35$ \citep{2006ApJ...638..686C}. 
This explains why only about one third of visually-classified major mergers and those with tidal tails in our sample are selected in Figure~\ref{fig:fig5}. 
We notice that some regular galaxies also show $A > 0.35$. It is mainly due to the contamination from neighboring sources within the target's 1.5 times Petrosian radius. A high $A$ is obtained for galaxy 51329 due to the warped disk as shown in Figure~\ref{fig:fig4}.

We emphasize that our $A_{\rm o}$ parameter focuses on the morphological disturbance in the OHR of a galaxy in comparison with $A$ that is dominated by the central high S/N light distribution and less sensitive to morphological disturbance in the OHR. It is clear that galaxies with tidal tails become largely indistinguishable from disks and irregular galaxies in Figure~\ref{fig:fig5}. Compared with Figure~\ref{fig:fig3}, we can conclude that our $A_{\rm o}$-$D_{\rm o}$ method is more efficient than $CAS$ in selecting galaxies with strongly disturbed morphologies like tidal tails.

\begin{figure}[] 
\centering
\includegraphics[width=0.48\textwidth]{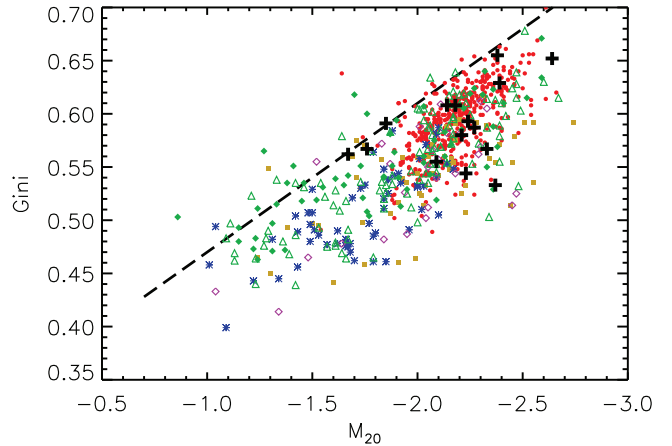}
\caption{$M_{20}$ versus Gini coefficient for our sample. The dashed line is the criterion ($G>-0.14\,M_{20}+0.33$) from \cite{2004AJ....128..163L} for selecting merging galaxies. All symbols are given as same as in Figure~\ref{fig:fig3}. The majority of regular galaxies lie below the dashed line and form a well defined sequence.}
\label{fig:fig6}
\end{figure}

\subsection{Gini-$M_{20}$}\label{sec:sec4.2}
      The Gini coefficient ($G$) was originally invented in economics and introduced to describe the light distribution of a galaxy through a Lorentz curve \citep{2003ApJ...588..218A,2004AJ....128..163L}. The Gini coefficient ranges from 0 to 1. A lower $G$ means a more uniform flux distribution within a galaxy while a higher $G$ means that a higher fraction of the total flux is contained in a smaller fraction of the galaxy pixels. $M_{20}$ is a parameter to measure the second-order moment of the light distribution of the top 20\% brightest pixels in a galaxy which is then normalized to the light moment for all pixels. Like parameter $C$, it describes the central concentration of the brightest pixels of the galaxy depending on their distance from the galaxy center. 

    Figure~\ref{fig:fig6} shows the distribution of our sample in the Gini-$M_{20}$ diagram. We can see that our sample galaxies at intermediate redshifts are still distributed along a tight sequence following the nearby galaxies given in \citet{2004AJ....128..163L}. The early-type galaxies have higher $G$ and lower $M_{20}$ (top-right) while late-type galaxies have lower $G$ and higher $M_{20}$ in a statistical sense. \citet{2004AJ....128..163L} suggested to select major mergers by the criterion $G>-0.14\,M_{20}+0.33$ (the dashed line in Figure~\ref{fig:fig6}). However, there are only 9 of 56 major mergers (green diamonds) and  one of 15 galaxies with tidal tails (black crosses) located above the dashed line. It has been pointed out that that mergers identified by the Gini-$M_{20}$ method are not complete \citep[e.g.,][]{2007ApJS..172..329K,2007ApJS..172..406S,2010ApJ...721...98K}. Simulations show that the Gini-$M_{20}$ method primarily enables identifying mergers during the first pass or final stage when the galactic nuclei are distinguishable \citep{2008MNRAS.391.1137L,2010MNRAS.404..575L,2010MNRAS.404..590L}. The simulations also uncover that Gini-$M_{20}$ is insensitive to mergers in the intermediate stage (i.e., maximal separation and/or second pass) when tidal tails may be formed. Comparison of Figure~\ref{fig:fig6} with Figure~\ref{fig:fig3} clearly demonstrates that our $A_{\rm o}$-$D_{\rm o}$ method is better than Gini-$M_{20}$ in identifying major mergers and galaxies with apparent extended tidal features. 

\section{DISCUSSION}\label{sec:sec5}

    Two parameters, outer asymmetry and centroid deviation, are developed to quantify the structure in the OHRs of galaxies, which usually have lower surface brightness than their IHRs. The tidal features like tidal tails in the OHRs are signatures of major mergers. Such delicate features are often extended and faint. Detecting low surface brightness emission in the OHRs is thus critical to the measurements of the two parameters. 

We adopt a low threshold (0.8\,$\sigma_{\rm bkg}$) for source detection with SExtractor in order to probe the extended faint structures around a target. 
This may mis-identify some surrounding noise pixels as part of the target, yielding a fuzzy appearance although the contribution to the total flux is negligible  (see Figure~\ref{fig:fig1}). The effect of noise on the structural properties in the OHR would become significant for low surface brightness galaxies. Such a detection technique based on surface brightness threshold has been widely used in studies of galaxy morphologies \citep[e.g.,][]{2003ApJ...588..218A}. However, the physical size of detection of a given galaxy relies on the depth of the imaging and redshift due to the cosmological dimming effect. Then the detection completeness of tidal features needs to be addressed for the evolution of a population of galaxies with dedicated features. \citet{2007ApJ...656....1L} adopted surface brightness threshold changing as $(1+z)^3$ to compensate for the cosmological dimming effect and ensure a constant surface brightness cut for all redshifts examined.  This is beyond the scope of this work. 

Alternatively, the elliptical or circular Petrosian radius \citep{1976ApJ...209L...1P} derived from the growth curve of a galaxy can be used to determine the segmentation map of the galaxy. Taking the pixels within the Petrosian radius as the segmentation map is able to correct the cosmological dimming effect \citep{2003ApJS..147....1C,2004AJ....128..163L}. Nevertheless, this treatment may sacrifice the detection sensitivity of extended tidal features in the OHRs. For instance, the typical tidal tails of merging galaxies can be as extended as several tens of kpcs \citep{2007ApJ...663..734E} and beyond roughly 1$\sim$1.5 times Petrosian radius of the host galaxies. 

\citet{2007ApJ...669..184A} developed the so-called quasi-Petrosian segmentation method and claimed that it enables the probe of extended low surface brightness structures and the correction for the cosmological dimming effect.  Firstly, all pixels of a galaxy in the SExtractor preliminary segmentation map are sorted in decreasing order of flux. The method will select pixels from the highest end of flux to the lowest end until a pixel meets $f_i=\eta(F_i/i)$, where $f_i$ is flux of pixel $i$, and $\eta$ is scale factor. $F_i/i$ is the cumulative mean surface brightness. However, the fuzzy sky pixels will still be included when $\eta$ is set lower for the detection of faint tidal tails.

\section{CONCLUSION}\label{sec:sec6}

We develop a new automatic method to quantify the structure in the outskirts of galaxies, aiming at probing delicate features like tidal tails. Using the isophote which encloses half the total light of a galaxy, the division of the galaxy image into two sections (the IHR and OHR) is the key to our method. Two parameters are introduced in the method: $A_{\rm o}$, which measures the asymmetry of the OHR, and $D_{\rm o}$, which measures the deviation of the intensity weighted centroids of the OHR from that of the IHR relative to the effective radius. The galaxies with stronger disturbance in morphology are expected to have higher $A_{\rm o}$ and $D_{\rm o}$. Moreover, the two parameters are designed to be less affected by the central high surface brightness section of galaxies, and thus sensitive to low surface brightness features in the OHR. A sample of 764 galaxies with $\log (M/{\rm M}_\odot)>3\times 10^{10}$ and $0.35<z<0.9$ selected from the GEMS and GOODS-S surveys is used to verify our method. For a comparison, we visually classify morphologies for the sample using $HST$ $z_{850}$-band images, following the usual classification scheme given in the literature. The $z_{850}$ band corresponds to the rest-frame optical over the redshift range examined here.

Our investigation shows that all sample galaxies fall on a sequence in the $A_{\rm o}$-$D_{\rm o}$ space. The position along the sequence is in general correlated with the degree of morphological disturbance. Galaxies with more disturbed morphologies have higher $A_{\rm o}$ and $D_{\rm o}$ in a statistical sense. The merging galaxies with tidal tails are well separated from regular galaxies (spheroids and disks) along the sequence. The regular galaxies are mostly with $D_{\rm o}<0.5$ and $A_{\rm o}<0.6$, following a relation described by ${\rm log}\,A_{\rm o}=0.6\,{\rm log}\,D_{\rm o}$. 
The galaxies with tidal tails are mostly with $A_{\rm o}>$0.5 and $D_{\rm o}$ ranging from 0.3 to 1.4. The criterion ${\rm log}\,A_{\rm o} > -1.6\,{\rm log}\,D_{\rm o}-1.1$ is able to select all galaxies with tidal tails and most major mergers (without apparent tidal tails). 

The advantage of this selection is that major mergers with highly asymmetric morphologies are nearly complete and about 87\% of galaxies with regular morphologies (spheroids and disks) are excluded at the same time. The left 13\% of regular galaxies tend to have higher $A_{\rm o}$ and $D_{\rm o}$ contributed by odd spiral arms, strong dust lanes, warped disks,  or contamination from neighboring sources.  Low surface brightness galaxies suffer more from the contamination by surrounding sources or background noise, leaving $A_{\rm o}$ and $D_{\rm o}$ highly uncertain. 
Given that nearly 63\% of the major mergers (including those with tidal tails) can be selected by a single cut in the $A_{\rm o}$-$D_{\rm o}$ space, out method can be used for a relatively complete search for major mergers from large scale imaging surveys. 

Compared with $CAS$ and Gini-$M_{20}$, our $A_{\rm o}$-$D_{\rm o}$ method is able to provide a better separation between mergers and regular galaxies. In particular, our method is unique in probing extended delicate features. In Gini-$M_{20}$, galaxies with tidal tails mix together with regular galaxies. And $CAS$ is also insensitive to galaxies of this kind. We point out that our $A_{\rm o}$ is in a better position than $A$ in $CAS$ for probing asymmetric structures of galaxies. 

Next-generation deep imaging surveys carried out by the Large Synoptic Survey Telescope (LSST) and Euclid will provide high-quality multi-band imaging data over large areas of sky, allowing for detailed studies of the structure and morphology of galaxies to address how galaxies form and evolve. Automatic approaches like our $A_{\rm o}$-$D_{\rm o}$ method for morphological analysis are essential in the era of big data.

\acknowledgments
We are grateful to the referee for the helpful suggestions that improved this paper. We also thank Yinghe Zhao, Yewei Mao and Cong Ma for stimulating discussions. This work is supported by National Basic Research Program of China (973 Program 2013CB834900) and the Strategic Priority Research Program "The Emergence of Cosmological Structures" of the Chinese Academy of Sciences (Grant No. XDB09000000).

\end{document}